\documentstyle[aps]{revtex}
\begin{document}
\centerline{\LARGE\bf Everyone can understand quantum mechanics}
\vskip 1cm \centerline{Gao Shan} \centerline{Institute of Quantum
Mechanics} \centerline{11-10, NO.10 Building, YueTan XiJie DongLi,
XiCheng District} \centerline{Beijing 100045, P.R.China}
\centerline{E-mail: gaoshan.iqm@263.net}

\vskip 1cm
\begin{abstract}
We show that everyone can understand quantum mechanics, only if he
rejects the following prejudice, namely classical continuous
motion (CCM) is the only possible and objective motion of
particles.

\end{abstract}

\vskip 1cm

{\it I think I can safely say that nobody today understands
quantum mechanics.    ------Feynman (1965)}

\vskip .5cm

When people talk about motion, they only refer to CCM, its
uniqueness is taken for granted absolutely but unconsciously,
people never dream of another different motion in Nature, but to
our surprise, as to whether or not CCM is the only possible and
objective motion, and whether CCM is the real motion or apparent
motion, no one has given a definite answer up to now.

In classical mechanics, CCM is undoubtedly the leading actor,
while in quantum mechanics, CCM is rejected by the orthodox
interpretation from stem to stern, but why did people never guess
what quantum mechanics describes is just another different motion
from CCM? as we think, this is the most direct and natural idea,
since classical mechanics describes CCM, then correspondingly
quantum mechanics will describe another kind of motion.

The only stumbling block is just the huge prejudice rooted in the
mind of people, it is that classical continuous motion (CCM) is
the only possible and objective motion of particles, now let's see
it more clearly through looking back to the history.

Bohr\cite{Bohr} and his enthusiastic supporters held this
prejudice strong, they insisted that Copenhagen interpretation is
the only possible interpretation of quantum mechanics, since CCM
can no longer account for the phenomena in quantum mechanics, we
must essentially discard it, the only possible and objective
motion, then it is evident that quantum mechanics provides no
objective description of Nature at all, but only our knowledge
about Nature.

Einstein\cite{Einstein} held this prejudice stronger, he believed
that if the objective picture of classical continuous motion
contradicts with quantum mechanics, the wrong side can only be
quantum mechanics, not classical continuous motion, since in any
case we can not lose the reality, while classical continuous
motion is the only reality of Nature, thus he became the strongest
opponent of Copenhagen interpretation, but his acerbic comments
did not help him so much, he failed in persuading Bohr, as well as
his contemporary.

Bohm\cite{Bohm} also held this prejudice, his cleverness lies in
that he provided a compromise hidden-variable picture between
those of Bohr and Einstein, but neither one was satisfied with his
way, and he himself was also tortured by the dualistic monster he
created.

Everett\cite{Everett} still held this prejudice, even though he
presented a crazy many worlds interpretation for quantum
mechanics, his interpretation is still in the framework of CCM,
only for every branch of the expensive many worlds, and no
supporters would like to attempt quantum suicide to convince
themselves the many worlds interpretation is right, let alone
convince anyone else.

More and more followers have been trying to understand quantum
mechanics, but they still held this prejudice firmly and
unconsciously, they are doomed to fail, this is their destiny due
to the prejudice.

Then why cling to it till death like a miser? unloosen it! please
reject it! and don't walk along this wrong way any more, it only
leads to the blind alley, the impasse, no way out there.

In our previous paper\cite{Gao}, from the clear logical and
physical analyses about motion, we have shown that the natural
motion in continuous space-time is not CCM, but one kind of
essentially discontinuous motion, and Schr\"{o}dinger equation in
quantum mechanics is just its simplest nonrelativistic motion
equation; while in the real discrete space-time, the natural
motion is also discontinuous, and it will result in the collapse
process of the wave function, this brings about the appearance of
CCM in macroscopic world, thus CCM is by no means the real motion
in Nature, let alone be the only possible and objective motion, it
is just one kind of ideal apparent motion in the macroscopic world
where we live, while the real motion is essentially discontinuous.

Once we reject the apparent CCM, and find the real motion in
Nature, understanding quantum mechanics is just an easy task, we
can safely say that everybody can understand quantum mechanics
easily from now on, nobody will be plagued by its weirdness any
more, since quantum mechanics is just the theory describing the
real motion in Nature, even though the real motion is more complex
than CCM, it also has a clear picture just like CCM, its weirdness
results only from its particular existence and evolution, in fact,
from a logical point of view, its existence and evolution are more
natural than those of CCM, only because we are unfamiliar with it,
it looks very bizarre for us.

Concretely speaking, the wave function $\psi(x,t)$ in quantum
mechanics is an indirect mathematical complex to describe the
state of the real motion of particle, the direct description
quantities are $\rho(x,t)$ and $j(x,t)$, their relation is
$\psi(x,t)=\rho^{1/2}\cdot{e^{iS(x,t)/\hbar}}$, where
$S(x,t)=m\int_{-\infty}^{x} j(x^{'},t)/\rho(x^{'},t)dx^{'}+C(t)$,
the apparent wave-like form of $\psi(x,t)$ results essentially
from the discontinuity of the real motion, not from any objective
existence of wave or field.

The evolution of the real motion includes two parts, one is the
linear evolution part, it results in the interference pattern,
which is usually the display of classical wave, but the pattern is
undoubtedly formed by a large number of particles undergoing the
real motion; the other is the nonlinear stochastic evolution part,
it results in the collapse process of the wave function, during
measurement this process happens very soon, and the wave function
of the particle collapses into a local region, this brings about
the appearance of single event in measurement, this process is
stochastic and indeterministic due to the essential discontinuity
and randomicity of the real motion itself.

Certainly, one point needs to be stressed, even though the wave
function does provide a complete description of the state of the
real motion, present quantum theory does not provide a complete
description of the evolution of the real motion, and needs to be
revised to include the stochastic evolution part.

Now we may also understand why people haven't understood quantum
mechanics yet after they found it more than seventy years ago, the
reason is very simple, because people always discuss and picturize
it in the framework of CCM, they can only see the sky of CCM, some
of them would ruthlessly reject the reality in the quantum world
rather than give another possible motion a glance, the others
would never ever give up CCM, this is indeed the sorriness of
science, but the most heart-struck is that people are always very
complacent about their own choices, and care little about the
ideas of others, all these will be fundamentally changed from now
on.

\end{document}